\documentclass{article}
\usepackage[T1]{fontenc} 
\usepackage[utf8]{inputenc} 
\usepackage{ismir,amsmath,cite,url}
\usepackage{graphicx}
\usepackage{color}
\usepackage{soul}
\usepackage{harmony}
\usepackage{listings}
\usepackage{multirow}
\usepackage{subcaption}

\title{A holistic approach to polyphonic music transcription with neural networks}

\threeauthors
  {Miguel A. Román} {U.I. for Computing Research \\ University of Alicante, Spain \\ {\tt mroman@dlsi.ua.es}}
  {Antonio Pertusa} {U.I. for Computing Research \\ University of Alicante, Spain \\ {\tt pertusa@dlsi.ua.es}}
  {Jorge Calvo-Zaragoza} {U.I. for Computing Research \\ University of Alicante, Spain \\ {\tt jcalvo@dlsi.ua.es}}

\begin{document}

\maketitle
\begin{abstract}
We present a framework based on neural networks to extract music scores directly from polyphonic audio in an end-to-end fashion. Most previous Automatic Music Transcription (AMT) methods seek a piano-roll representation of the pitches, that can be further transformed into a score by incorporating tempo estimation, beat tracking, key estimation or rhythm quantization. Unlike these methods, our approach generates music notation directly from the input audio in a single stage. For this, we use a Convolutional Recurrent Neural Network (CRNN) with Connectionist Temporal Classification (CTC) loss function which does not require annotated alignments of audio frames with the score rhythmic information. We trained our model using as input Haydn, Mozart, and Beethoven string quartets and Bach chorales synthesized with different tempos and expressive performances. The output is a textual representation of four-voice music scores based on **kern format. Although the proposed approach is evaluated in a simplified scenario, results show that this model can learn to transcribe scores directly from audio signals, opening a promising avenue towards complete AMT.  
\end{abstract}
\section{Introduction}\label{sec:introduction}

Automatic music transcription (AMT) aims to convert acoustic music signals into any sort of music notation. Most of the music we listen today is polyphonic, where simultaneous sound events produced by different audio sources (i.e., instruments) are combined in a single acoustic waveform. This aggregation process entails loss of information, making the transcription task very challenging even for trained musicians. Moreover, the different sound events are highly correlated in time and frequency due to the rhythmic and harmonic patterns usually found in music, which complicates sound separation even further as we cannot rely on the statistical independence of the source signals. Therefore, in order to produce a proper music score from an audio signal, multiple complex sub-tasks must be involved such as multi-pitch estimation, note onset/offset detection, source separation, as well as other musical context information retrieval tasks like metering and tonality estimation. 

As pointed out by \cite{Benetos:19}, there are many approaches to tackle AMT, yet most works focus on solving only one intermediate goal of the whole problem. Frame-level transcription, also known as multi-pitch estimation, aims to detect which fundamental frequencies are present at each time step of the input signal. Note-level transcription goes a step further by estimating the notes characterized by their pitch and clock-time duration (onset and offset times), producing a piano-roll representation of the music. Stream-level transcription extends the note-level approach by associating each note with its originating instrument based on its timbre. Lastly, the notation-level transcription is the final goal of AMT, aiming to produce a music score with enough information to interpret the original recording. 

In this work, we denote the notation-level transcription as Audio-to-Score (A2S) task, where the audio signal is processed to be converted into a symbolic music score. Even with a perfect transcription, the output of any A2S system cannot faithfully represent the music that was originally played. It must be considered that musical audio signals are often expressive performances, rather than simple mechanical translations of notes read from a staff. A particular score can be performed by a musician in many different ways, and similarly there are several ways to represent the same musical excerpt with standard music notation (e.g., a dotted half note is ``the same'' as a half note tied to a quarter note). Music scores can only be seen as guides to aid musicians, highly correlating but never fully explaining musical experience. This makes A2S a rather ill-defined problem without unique solutions.

Despite the above, our work aims to demonstrate that the A2S task can be performed in a single step. To this end, we make use of a deep neural network that is trained in an  end-to-end fashion to produce a sequence of musical symbols that describes a feasible polyphonic score out of the input audio. Our experiments are conducted using Haydn, Mozart, and Beethoven string quartets and Bach chorales synthesized with different tempos and expressive performances.\footnote{The source code and data are available at \url{https://github.com/mangelroman/audio2score}.} Although the analysis of the current performance requires a deeper reasoning regarding evaluation metrics, we provide some results that account for the good performance of the proposed model and allow us to be optimistic about this line of research.

\subsection{Related Work}\label{sec:relatedwork}

There are recent AMT approaches using deep neural networks for the multi-pitch detection task \cite{kelz:17, kelz:16, magentaismir18}. For this, Short-Term Fourier Transform (STFT), log-frequency STFT or Mel spectrograms are usually fed to Convolutional Neural Networks (CNN) to extract piano-roll representations as output. Other works focus on producing music scores from unquantized MIDI representation \cite{Cogliati:16}.

One of the few methods aiming to extract a complete score directly from audio is that of \cite{Nakamura:18}. In this work, a multi-pitch detection method with note tracking is used to get a piano-roll representation that is further converted into a quantized MIDI file by using a rhythm quantization method \cite{nakamura17}. Afterwards, a score typesetting software such as MuseScore can be used to get a MusicXML file from the MIDI output.

To the best of our knowledge, there are only two approaches that perform A2S in a single stage, directly converting the input audio into any music notation format. This has the advantage that a wrong detection in a given stage (such as the multi-pitch detection) is not propagated through the next processing stages, avoiding error cascading. The only works addressing notation-level AMT in an end-to-end manner are those of \cite{Carvalho:17} and \cite{Roman:18}. Both works follow a supervised learning approach with deep neural networks to solve the AMT task in one step. Although they bring promising results, the proposed models include several limitations (e.g. monophonic audio in \cite{Roman:18} and fixed input length in \cite{Carvalho:17}) that cannot be disregarded when addressing the notation-level AMT problem as a whole. 

In \cite{Carvalho:17}, authors show how a Convolutional-Recurrent Neural Network architecture (CRNN) \cite{Shi:17} can learn all the basic tasks involved in notation-level AMT, but it is only a very limited proof of concept that cannot address most of the possible scores. In the second of these works \cite{Roman:18}, the AMT problem was addressed as an Automatic Speech Recognition (ASR) problem. By using monophonic audio as input and a sequence of symbols (analogously to the written language characters) as output, several methods that were originally developed for ASR can be used. In particular, \cite{Roman:18} adopted an architecture inspired in DeepSpeech2 \cite{deepspeech2}, which learns to map audio frames to a sequence of characters without any alignment.

Training with unaligned data, i.e. without needing the input audio frames to be aligned with the music symbols, is a clear advantage as much more data can be gathered without going through the tedious task of manually annotating the location of the output symbols in their corresponding input audio frames. Nevertheless, monophonic audio transcription does not exhibit the essential challenge coming from simultaneous sound events. The present work goes one step beyond by showing that a similar formulation can also be reliable for polyphonic music.

\section{Data} \label{sec:data}

\begin{figure}
 \centerline{
 \includegraphics[width=\linewidth]{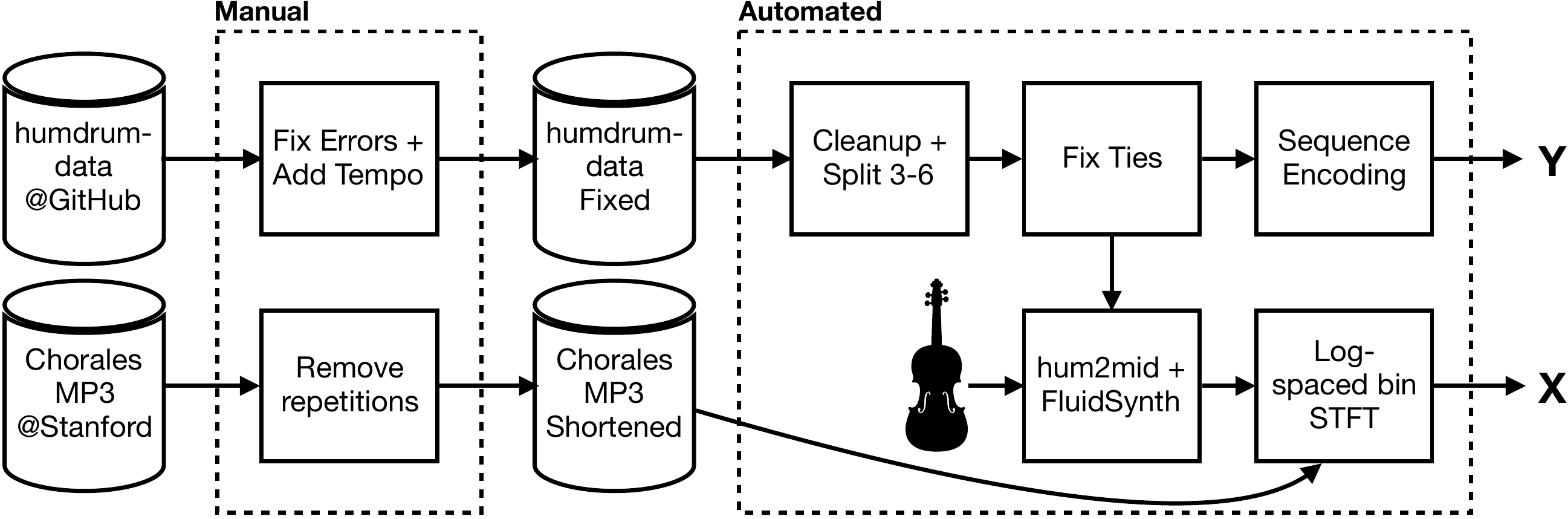}}
 \caption{Data acquisition pipeline showing the manual and automated steps required to build the ground truth.}
 \label{fig:pipeline}
\end{figure}

\begin{figure}[t]
 \includegraphics[width=\columnwidth]{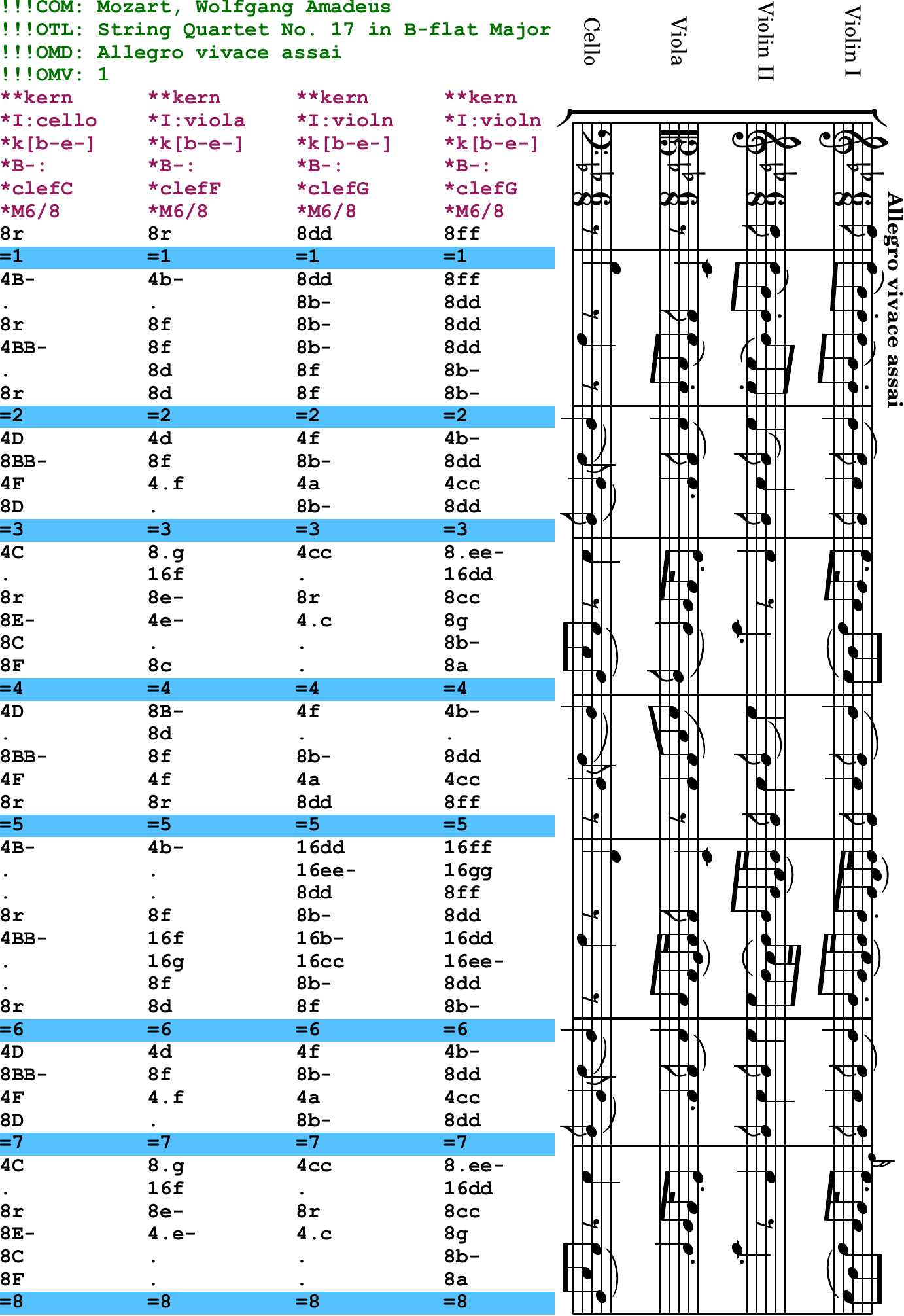} 
 \caption{Example of one score in **kern format (left) representing the output of our model and the rendered western music score (right).}
 \label{fig:kernexample}
\end{figure}

Our notation-level AMT approach, namely A2S, seeks to estimate which music score, modeled as a structure containing symbols from a fixed alphabet of music notation, would likely define that audio. 

Let $\mathcal{X}$ be the domain of audio files and $\Sigma$ the alphabet of music score symbols. The aim of our A2S is to compute a function that maps any audio file into a sequence of symbols, i.e., a function $f : \mathcal{X} \rightarrow \Sigma^{*}$. 

\subsection{Input Representation}\label{subsec:input}
The input representation of our model is the spectral information of the raw audio waveform over time, based on the STFT with log-spaced bins and log-scaled magnitude. In this type of spectrogram, frequency bins are aligned with equal-tempered music scales using 440Hz as the reference for A4 pitch. The sampling rate of the input audio files was 22,050Hz, and STFT was calculated with a Hamming window with size 92.88ms (2048 samples) and a hop of 23.22ms (512 samples). Only frequencies between pitches C2 and C7 were considered, extracting 48 bins per octave. 

\subsection{Output Representation}\label{subsec:output}
The output music notation of our model is a single sequence of symbols that can be used to render a multi-part western music score. These symbols represent both notes and rests with their corresponding duration, barlines, ties between notes, and fermatas. It is important to remark that in the context of the A2S task, notes are not the same as pitches. For example, pitch 349.23Hz can be represented as F4, E$\sharp$4 or G$\flat\flat$4 depending on the key signature. 

We are not including clefs in our output representation, as they are only intended to aid in the score visualization and do not carry any musical information we can extract from the audio. For the sake of simplicity, we are also not including time signatures in the output sequences assuming they can be inferred from the predicted barlines for the type of scores in our training set. By the same rationale, key signatures are neither included assuming they can also be inferred from the predicted notes. Moreover, since most of our samples are made of fragments of much longer scores, they may not carry enough information to predict the correct key signature, therefore misleading the training process. Other music notation symbols such as slurs, grace notes, ornaments, and articulations marks are also left out of the scope of this work.

\subsection{Data Preparation}\label{subsec:data}

As previously mentioned, the fact that we do not require data alignments is a clear advantage to easily build the ground truth needed to train our model. However, the majority of scores available in the public domain are usually in printed form, so we cannot automatically obtain the symbolic representation we need unless we make use of an Optical Music Recognition system. A lot of progress has been made in this area of study but unfortunately it is still insufficient to meet our precision needs, driving us to look for existing text based scores instead. After some analysis of the various types of music encoding formats within reach, we chose the humdrum toolkit \cite{humdrum} due to its versatility to represent polyphonic music.

The humdrum file format is a general-purpose human-readable 2D representation of music information intended to assist music researchers. The columns of the text file, separated by the tab character, represent the sources of information that produce music-related events. The lines of the text file represent the evolution of those events over time. The humdrum syntax defines the skeleton that contains other higher level schemes of music notation, like the **kern format our ground truth is based on. The **kern format is designed to encode the semantics of a western musical score, rather than the visual aspects of its printed realization, matching nicely with the purpose of this work.

An example of a music excerpt encoded in **kern notation is shown in \figref{fig:kernexample} along with its associated sheet music excerpt. In this format, columns are called spines and they are associated with instruments, just like a pentagram in western sheet music. Spines may contain one single sound event or the combination of various sound events with the same canonical duration, namely a chord. Spines can also be split into two spines when two independent voices (excluding chords) occur for the same instrument. The newly created spine can be rejoined back to the original spine when the extra voice is no longer needed. This level of flexibility gives almost no restrictions to the kind of music it can support, making **kern a good candidate to endure future work. 

We created two datasets out of the **kern files available in the humdrum-data repository \cite{humdrumdata}: the chorales dataset, containing 370 chorales of Bach, and the quartets dataset, containing most of the string quartets of Haydn, Mozart and Beethoven. In the chorales dataset we take each chorale as one training sample, and we use audio from expressive MIDI files synthesized with a high quality pipe organ soundfont \cite{stanfordchorales}. As we did not synthesize the audio, we had to manually remove repetitions to ensure that samples are not unnecessarily long. In the quartets dataset we randomly split the scores in fragments of 3-6 measures each, and we synthesized the corresponding MIDI file obtained from the hum2mid tool, which converts **kern to MIDI using dynamic spines and articulation marks when available in the original **kern file. We removed grace notes and ornaments from the score as they cannot be properly synthesized. We also removed split spines and upper notes of all chords to ensure no more than 4 simultaneous voices were present at any given time. Samples with double dots, double sharps or double flats are out of the scope of this work and therefore discarded. On the training set only, we allow overlapping of fragments as a means of data augmentation technique. \tabref{tab:datasets} summarizes the main characteristics of both datasets used in this work.

\begin{table}[t]
 \begin{center}
 \begin{footnotesize}
 \begin{tabular}{|l|l|l|}
  \hline
  & \textbf{Chorales} & \textbf{Quartets} \\
  \hline
  Number of samples  & 352 & 34,512 \\
  \hline
  Total duration  & 5.79h & 20.25h \\
  \hline
  Max duration  & 120s & 30s \\
  \hline
  Data Augmentation  & No & Yes \\
  \hline
  Polyphony voices  & 4 & 4 \\
  \hline
  \multirow{4}{*}{Instruments} & Pipe organ & Cello \\
  && Viola \\
  && Violin \\
  && Flute \\
  \hline
  Pitch range  & C2-A5 & C2-E7  \\
  \hline
  Shortest note  & $1/16^{th}$ & $1/64^{th}$ \\
  \hline
  Irregular groups  & None & Triplets \\
  \hline
  Tempo  & $\Vier\approx[60,70]$ & $\Vier=[40,200]$ \\
  \hline
  Vocabulary Size  & 99 & 143 \\
  \hline
  Train-test split \%  & 80/20 & 70/30 \\
  \hline
  Batch size & 4 & 16 \\
  \hline
 \end{tabular}
 \end{footnotesize}
\end{center}
 \caption{Summary of the datasets' characteristics.}
 \label{tab:datasets}
\end{table}
\figref{fig:pipeline} depicts the data acquisition pipeline we implemented to build our ground truth. The major inconvenience was dealing with multiple errors present in the **kern files, such as invalid ties, wrong canonical duration of notes and rests, and missing metronome markings. While these errors do not prevent musical analysis of the scores, they become noisy labels that hinder our training process. For that reason, we had to revise all the scores manually to correct these errors and label the missing metronome markings, with the help of existing tempo annotations and the conversion shown in \tabref{tab:tempos}. Adding metronome markings ensured the synthesized audio perform at reasonable speeds according to the composer's intention. Additionally, a random scaling factor in the $\pm6\%$ range was applied to each metronome marking to ensure tempo variability in all training samples.

\begin{table}[t]
 \begin{center}
 \begin{footnotesize}
 \begin{tabular}{|l|l||l|l|}
  \hline Largo Assai & 40 & Allegro Moderato & 120 \\
  \hline Largo & 50 & Poco Allegro & 124 \\
  \hline Poco Largo & 60 & Allegro & 130 \\
  \hline Adagio & 71 & Molto Allegro & 134 \\
  \hline Poco Adagio & 76 & Allegro Assai & 138 \\
  \hline Andante & 92 & Vivace & 150 \\
  \hline Andantino & 100 & Allegro Vivace & 160 \\
  \hline Menuetto & 112 & Allegro Vivace Assai & 170 \\
  \hline Moderato & 114 & Poco Presto & 180 \\
  \hline Poco Allegretto & 116 & Presto & 186 \\
  \hline Allegretto & 118 & Presto Assai & 200 \\
  \hline
 \end{tabular}
 \end{footnotesize}
\end{center}
 \caption{List of metronome markings chosen for classical music tempo annotations, given in number of quarter notes per minute.}
 \label{tab:tempos}
\end{table}
The resulting **kern scores after the preprocessing stage were then encoded in a special symbolic notation intended to reduce the number of characters and ease the training process. Accordingly, each canonical duration for notes and rests including their dotted version were encoded with just one symbol of the vocabulary. Likewise, note pitches were also encoded with one symbol condensing name and octave. In our **kern dataset, barlines are always repeated for all spines, so only one barline is maintained in the output representation referring to all spines. The rest of characters are preserved in the output representation in the same way, i.e. tabs, new lines, tie symbols, fermatas and the ``dot'' character, which indicates that the previous note/rest still affects the current row.

\section{Method}\label{sec:method}
Once the input and output representations are defined, we can formulate the A2S task as retrieving the most likely sequence of score symbols $\hat{\mathbf{s}}$ given an audio file $\mathbf{x} \in \mathcal{X}$:

\begin{equation} \label{eq:map}
\hat{\mathbf{s}} = \arg\max_{\mathbf{s} \in \Sigma^{*}} P(\mathbf{s}|\mathbf{x})
\end{equation}

\noindent where $\Sigma$ represents the set of characters necessary to encode the output as explained in the previous section (for instance, including ``tab'', ``new-line'' and ``dot'', among others). Additionally, $\Sigma$ includes an \emph{``empty''} symbol, denoted by $\epsilon$, that is necessary to separate two or more instances of the same symbol that occur in consecutive frames. 

Following successful approaches in other pattern recognition duties of similar formulation, we address this A2S with a holistic approach based on statistical models. Specifically, for learning the posterior probability provided in Eq.~\ref{eq:map}, we resort to Convolutional Recurrent Neural Networks (CRNN).

A CRNN is composed of one block of \emph{convolutional} layers followed by another block of \emph{recurrent} layers \cite{Shi:17}. The convolutional block is in charge of learning how to extract relevant features from the input and the recurrent layers interpret these features in terms of sequences of musical symbols. The activations in the last convolutional layer can be seen as a sequence of feature vectors representing the input audio file, $\mathbf{x}$.  Let $W$ be the width (number of frames) of the input sequence $\mathbf{x}$. The length of the resulting features after the convolutional layer will be $L =\gamma W$, where $\gamma\leq 1$ is implicitly defined by the specific configuration of the convolutional block (which usually includes some type of down-sampling to reduce dimensionality).

The output activations of the convolutional block are then fed to the first layer of the recurrent block, and the activations of its last layer can be considered proper estimates of the posterior probabilities per frame:

\begin{equation}
P(\sigma\mid \mathbf{x},j), ~ ~ 1 \leq l \leq L, ~ ~ \sigma \in \Sigma 
\end{equation}

\subsection{Training}
Convolutional neural networks can be trained through gradient descent using the well-known \emph{Back Propagation} algorithm. RNN networks can be trained similarly by means of \emph{Back Propagation Through Time}~\cite{Williams:95}. Therefore both the convolutional and recurrent blocks of a CRNN can be jointly trained by providing audio files annotated at the frame level.

In this work, however, we follow a holistic or ``end-to-end'' approach, which means that for each audio file we only provide its corresponding target transcript into score symbols, without any kind of explicit information about its segmentation into frames. A CRNN can be uniformly trained without this information by using the so-called Connectionist Temporal Classification (CTC) loss function~\cite{Graves:06}. The CTC training procedure is a form of Expectation-Maximization, similar to the backward-forward algorithm used for HMM training~\cite{rabiner:93}, that distributes the loss among all the frames to maximize Eq. \ref{eq:map} with respect to the ground-truth sequence. 

\subsection{Decoding}
In order to solve Eq. \ref{eq:map}, the most likely symbol is computed for each input feature vector of the recurrent block $l$, also referred as greedy decoding:
\begin{equation}
  \hat{\sigma}_l ~=~ \arg\max_{\sigma \in \Sigma}
    P(\sigma\mid \mathbf{x},l), ~ ~ 1\leq l\leq L
\end{equation}

Then, a pseudo-optimal sequence of musical symbols is obtained as $\hat{\mathbf{s}}\approx\mathcal{D}(\hat{\mathbf{\sigma}})$, where $\hat{\mathbf{\sigma}}=\hat{\sigma}_1\dots\hat{\sigma}_L$ and $\mathcal{D}\!:\Sigma^\star\rightarrow\Sigma^\star$ is a function which first merges all the consecutive frames with equal symbol, and then deletes all ``empty'' symbols~\cite{Graves:06}.


\subsection{Architecture}\label{subsec:architecture}

\begin{figure}
 \centerline{
 \includegraphics[width=\columnwidth]{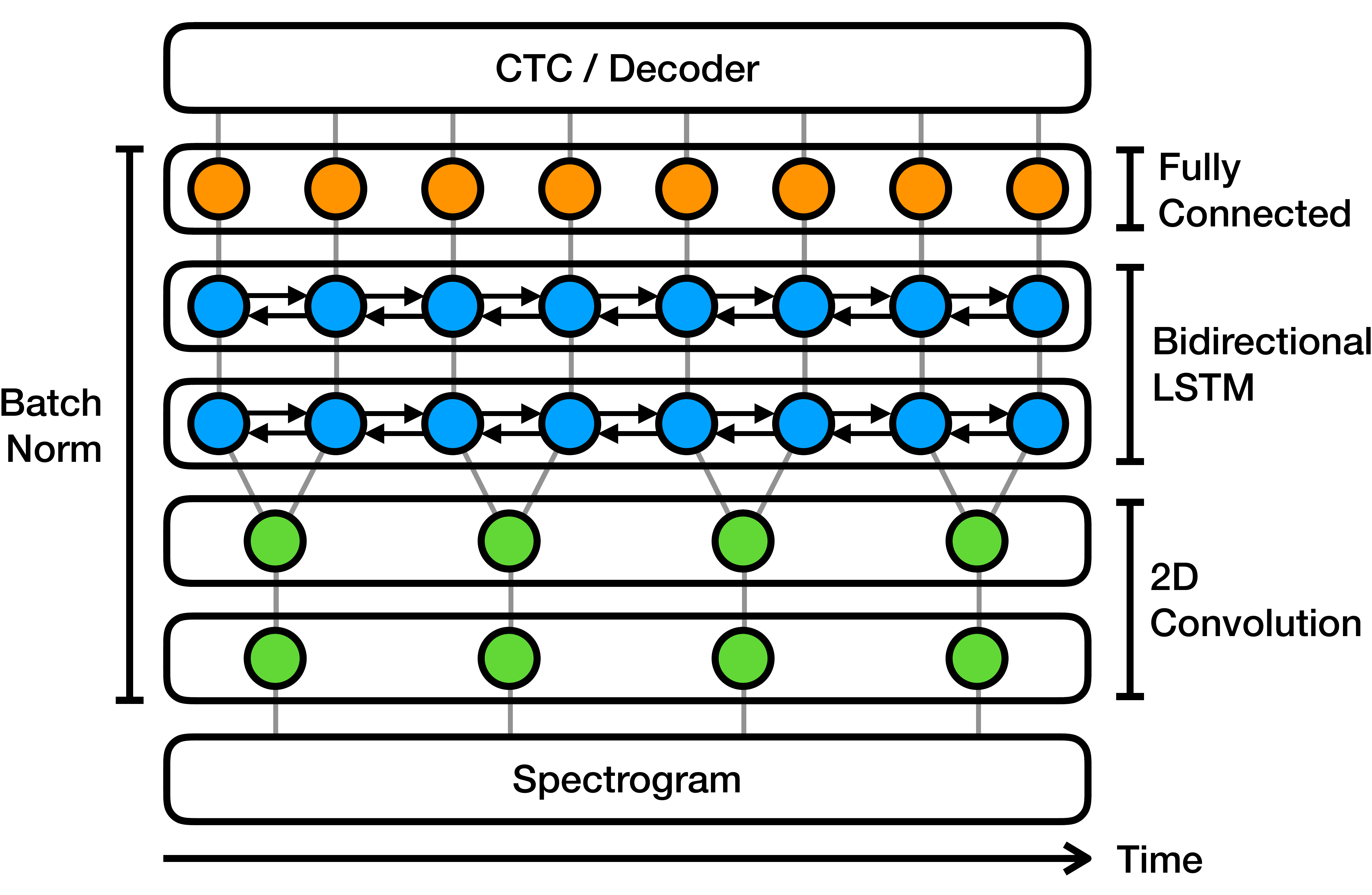}}
 \caption{High-level architecture of the Convolutional-Recurrent Neural Network used in our experiments.}
 \label{fig:architecture}
\end{figure}

The main building blocks of the CRNN considered for our experiments is illustrated in \figref{fig:architecture}.

The first two convolutional layers receive a 2D array containing the audio spectrogram described in section \ref{subsec:input} and apply $16$ filters of $3 \times 3$ with a stride of 2 in the frequency axis. We use filter striding to reduce the input dimensionality without the need of pooling layers.

For the Quartets dataset, output frames from the convolutional block are split in half, effectively doubling the number of frames feeding the next recurrent block. This is necessary to comply with the CTC loss function pre-condition for which the number of input frames must be greater or equal to the number of output symbols. Considering the high number of symbols per second in our sequence-based representation of a polyphonic score of the Quartets dataset, we apply this frame doubling technique at a lesser computational cost than increasing the density of the input spectogram.

The next two recurrent layers are based on Bidirectional Long Short-Term Memory (LSTM) cells, with $1024$ hidden units each. The fully connected layer at the end of the recurrent block converts the output per-frame predictions to the size of the output representation vocabulary.

With the purpose of reducing overfitting, Batch Normalization layers \cite{Ioffe:15} are added between any other layer excluding the input and output layers, as well as Dropout layers \cite{Srivastava:14} added after all the convolutional layers and after the last recurrent layer, with a drop probability of $0.1$ for the Quartets dataset and $0.2$ for the Chorales dataset. A higher drop probability is required for the Chorales dataset since less data is available for training, which increases the risk of overfitting.

\section{Experiments}\label{sec:experiments}

To the best of our knowledge, there are few specific evaluation metrics to measure the performance of a notation-level AMT method. In \cite{Nakamura:18} an evaluation metric for note-level AMT is discussed, but it still cannot be directly applied to our task (e.g. we do not have note onsets and offsets). \cite{Roman:18} adapts this metric to the A2S task by defining note duration errors instead of onsets/offsets errors. However, we believe this metric is still insufficient to properly evaluate a notation-level AMT method since, for instance, it does not take into account barlines and their effect in subsequent predictions of note durations and ties. The MV2H metric (Multi-pitch detection, Voice separation, Metrical alignment, note Value detection, and Harmonic analysis) was introduced in \cite{McLeod:18a}. This metric is closer to our needs, although its source code uses timing information in seconds that is not provided by our method. We leave it as an open point for future work to establish a proper notation-level AMT metric. 

In order to validate our method accuracy during training, we adopt the evaluation metrics from the ASR task as in \cite{Roman:18}, namely Word Error Rate (WER) and Character Error Rate (CER). They are defined as the number of elementary editing operations (insertion, deletion, or substitution) needed to convert the predicted sequences into the ground-truth sequences, at the word and character level respectively. Even though WER and CER are not specific to AMT, they provide a good indication of how close our score is to the ground-truth score.

In the context of our A2S task, we define words as any group of characters representing notes (including ties), rests and barlines in the output score. The ``tab'' and ``new line'' characters act as word separators, and only contribute to the CER calculation.

\subsection{Training process}\label{subsec:training}

\begin{figure*}
\centering
\begin{subfigure}[b]{0.45\linewidth}
 \includegraphics[width=\linewidth]{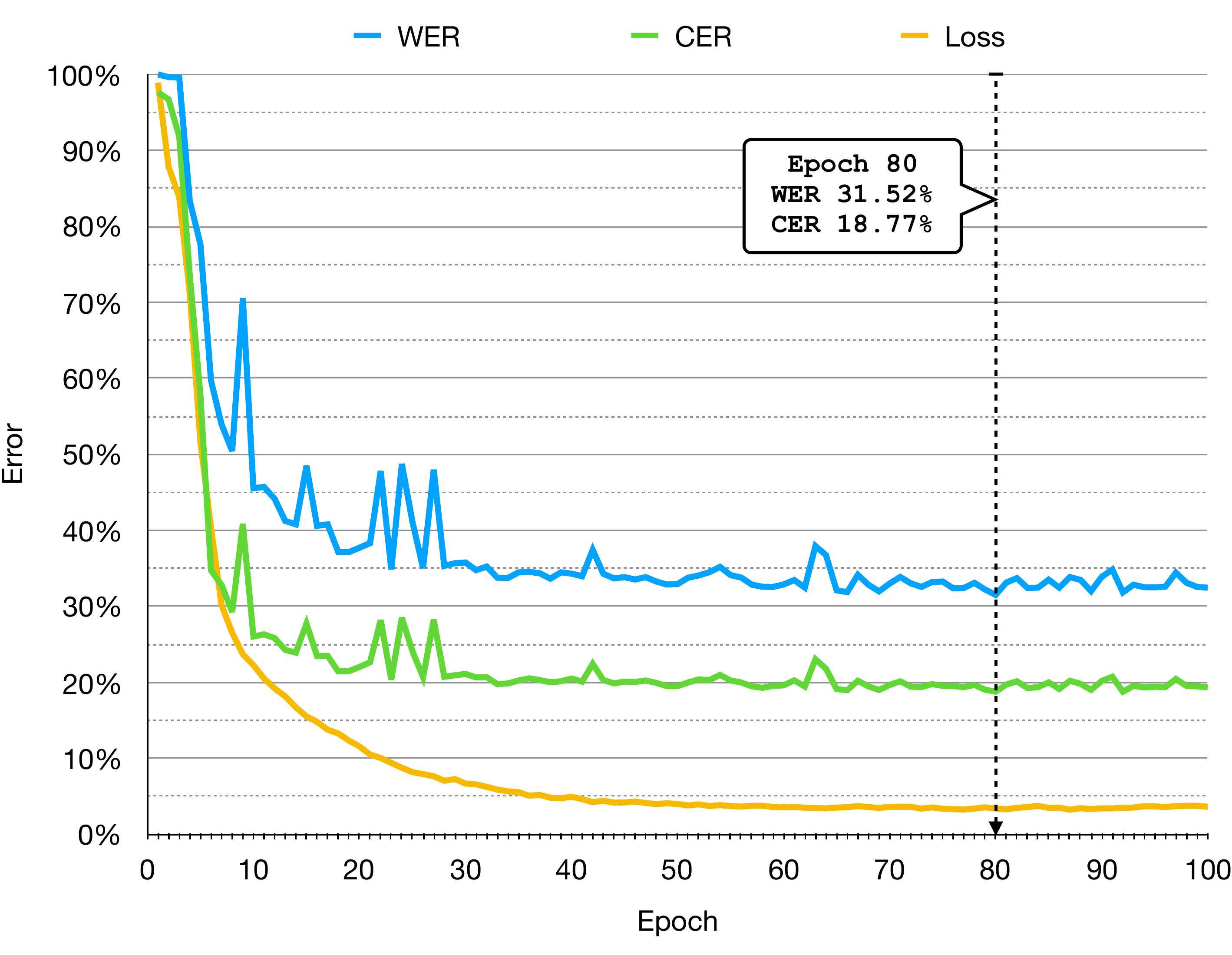}
 \caption{Chorales dataset.}
 \label{fig:trainingchorales}
\end{subfigure}
\begin{subfigure}[b]{0.45\linewidth}  
 \includegraphics[width=\linewidth]{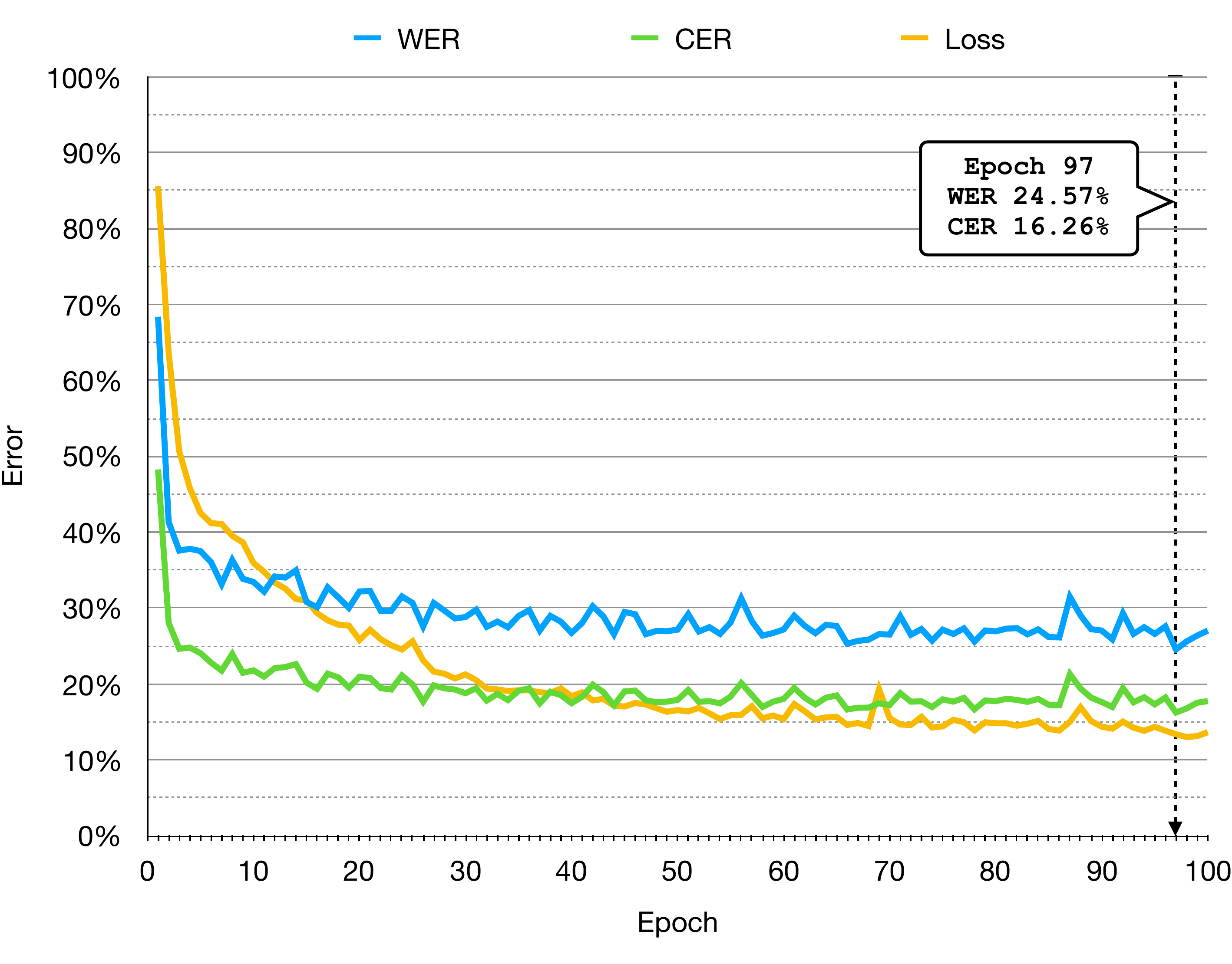}
 \caption{Quartets dataset.}
 \label{fig:trainingquartets}
 \end{subfigure}
 \caption{Evolution of loss, validation WER and CER during 100 epochs of training with a) Chorales dataset and b) Quartets dataset. Chorales WER is 30.96\% and CER is 18.10\%. Quartets WER is 18.10\% and CER is 13.53\%.}
 \label{fig:training}
\end{figure*}

\begin{figure*}[h!]
\centering
\begin{subfigure}[b]{0.4\linewidth}
 \includegraphics[width=\linewidth]{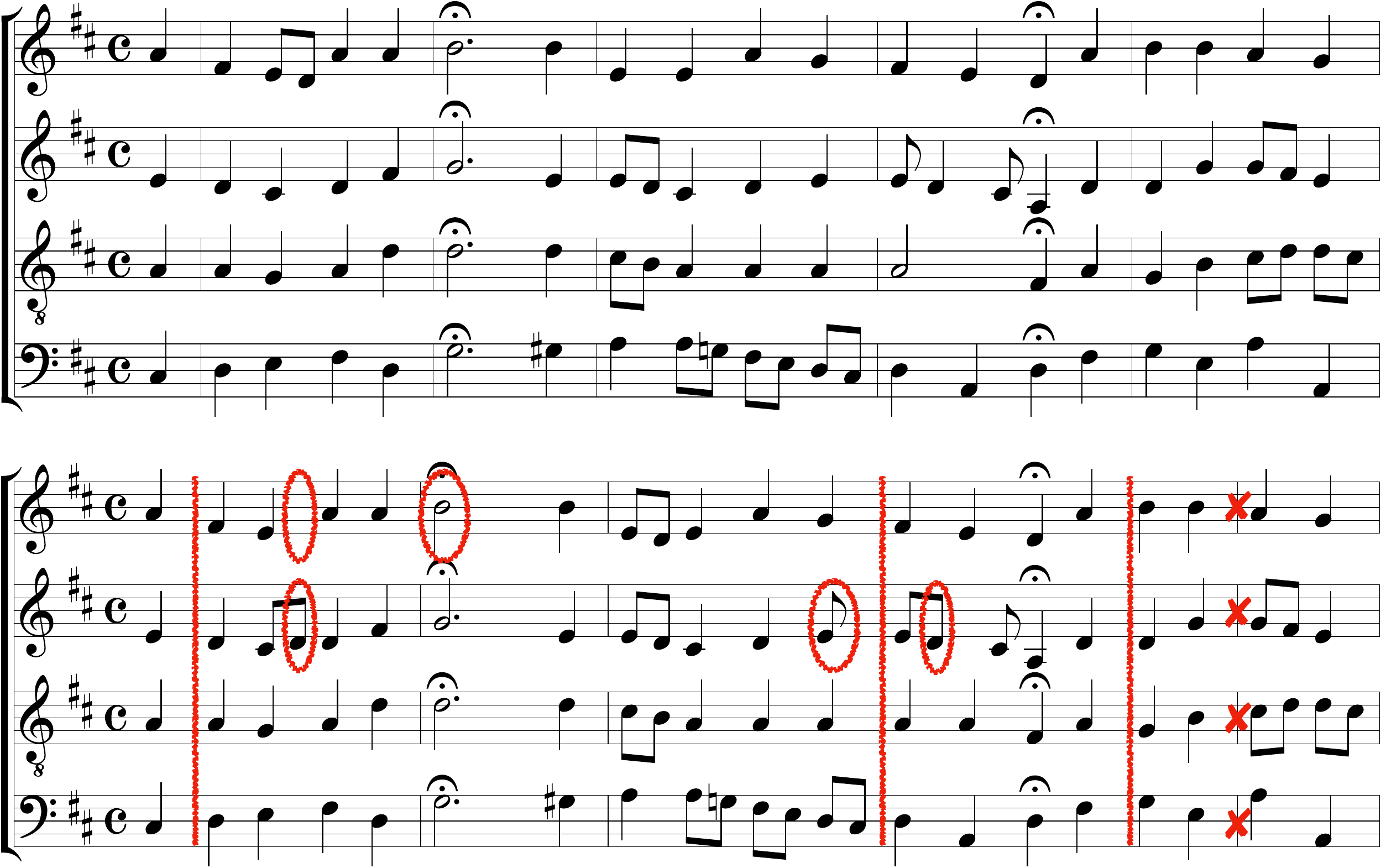}
 \caption{Chorales dataset example}
 \label{fig:choraleserrors}
\end{subfigure}
\begin{subfigure}[b]{0.45\linewidth}
 \includegraphics[width=\linewidth]{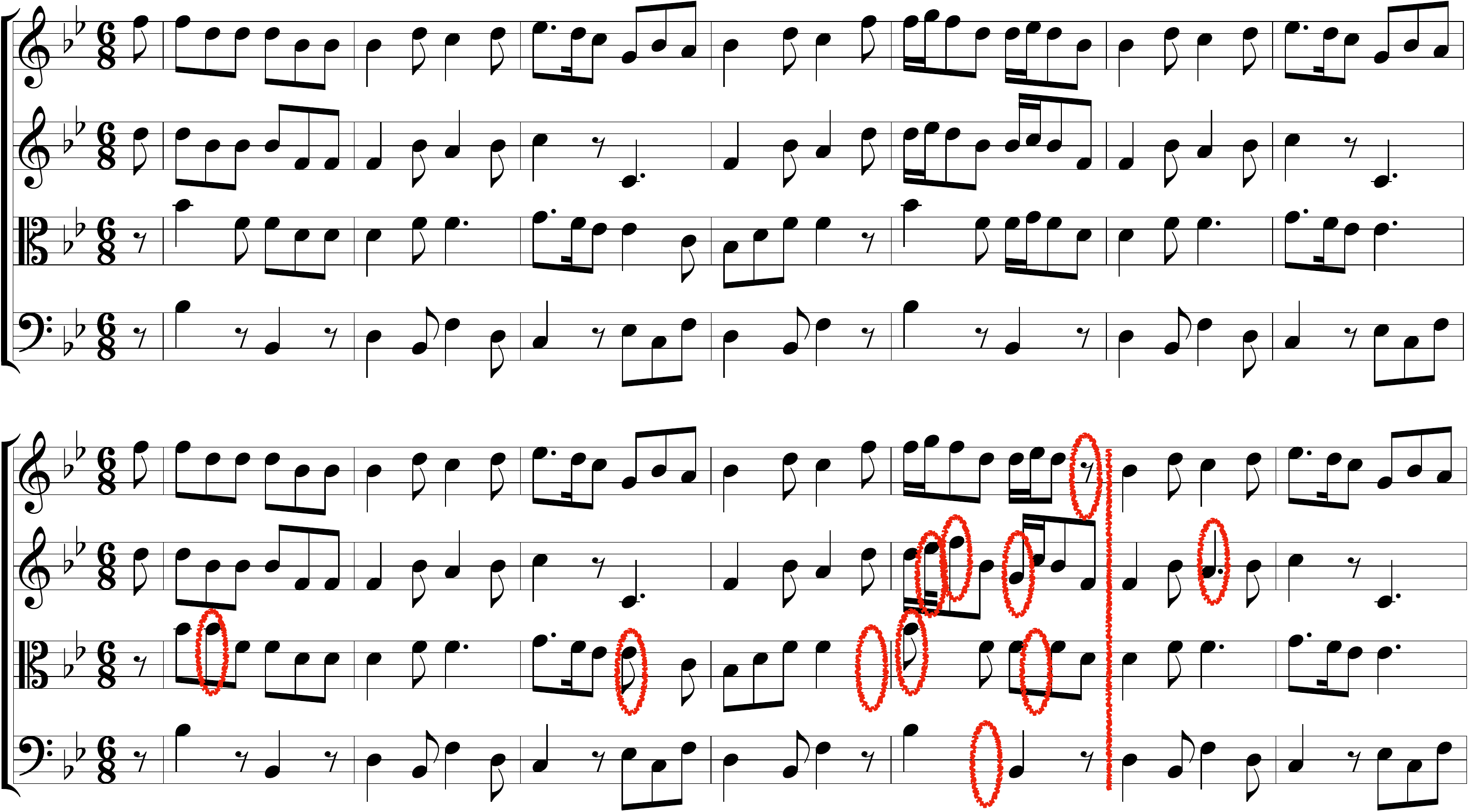}
 \caption{Quartets dataset example}
 \label{fig:quartetserrors}
 \end{subfigure}
  \caption{Excerpt of original (top row) and predicted scores (bottom row) from a test sample in a) Chorales dataset, b) Quartets dataset. The differences between original and prediction are highlight in red.}
  \label{fig:errors}
\end{figure*}

The models were trained for 100 epochs using mini-batch Stochastic Gradient Descent (SGD) optimizer, with Nesterov momentum of $0.9$. Our learning rate scheduling consists of 2 cycles of 50 epochs each, starting at $0.0003$ and annealing by $1.1$ at every epoch. After each epoch, the WER and CER are calculated for the validation set, and the model with the lowest WER is appointed as the best model for testing purposes.

The Chorales dataset, whose samples are full-length chorales, is trained with a batch size of 4. The Quartets dataset, whose samples are small excerpts extracted from the full-length quartets, is trained with a batch size of 16.  \figref{fig:training} shows the evolution of the CTC loss, WER and CER at training time on both datasets. Each figure also highlights the epoch where the best model was obtained.

\subsection{Results}\label{sec:results}
The best model obtained after the training process is then evaluated against the test set for both the Chorales and Quartets datasets, giving a WER of 30.96\% and CER of 18.10\% for Chorales, and WER of 21.02\% and CER of 13.53\% for Quartets.

After analyzing all test predictions, we observe that the model occasionally generates sequences that do not comply with the **kern format. Nevertheless, we believe these formatting errors can be solved by providing more samples to the training set or imposing syntax constraints. As shown in \figref{fig:errors}, most of the errors arise from wrongly estimated note durations and barlines. Exchanging notes between voices is another common mistake our model makes, specially when voice pitches are too close or even when two voices cross their melodic lines. 

The model struggles at predicting ties and triplets, which requires further analysis to determine whether it is related to barline errors, to the output representation format based on **kern, or to the lack of enough samples in the training set (i.e., ties and triplets are very infrequent in our training data compared to other symbols).

\section{Conclusions}\label{sec:conclusions}
In this work, we focus on the A2S task, a hardly explored formulation consisting of extracting a full score from an audio file. Note that A2S resembles what a human would expect to get if it intends to visualize the input audio as a music score (e.g., MusicXML), unlike what most authors consider AMT where the output sequence format is intended to be further processed by a computer (e.g., MIDI).

The proposed methodology which performs the A2S task has the following advantages over other AMT methods: 1) Frame-level alignment of the ground truth is not needed; 2) The end-to-end approach avoids propagating errors from one stage to the other; 3) The output of our model is based on **kern format and can be straightforwardly translated to a valid music score. 

We are aware this simplified scenario used for evaluation does not include real audio and some score symbols, but we argue the results provide the basis to open a new path of research towards notation-level AMT.

One of the main limitations of the proposed approach is the maximum-length of input sequences due to memory constraints. For example, this prevents an end-to-end training with complete songs, only allowing fragments of 2 minutes at a maximum on a typical training infrastructure. For this, we plan in a future work to explore other architectures such as the Transformer XL \cite{transformerXL}, a sequence-to-sequence model that can deal with much longer sequences. Other future works include defining an evaluation metric for A2S and building a dataset from real audio to validate the approach with actual music performances.

\section{Acknowledgment}
This work was funded by the Spanish Ministerio de Economía, Industria y Competitividad through HISPAMUS project (TIN2017-86576-R).


\bibliography{ISMIRtemplate}

%
%
%
%
\end{document}